\documentclass[preprint,12pt]{elsarticle}
\usepackage{amsmath,amssymb}
\usepackage{booktabs}
\usepackage{graphicx}
\usepackage{hyperref}

\begin{document}

\begin{frontmatter}

\title{Computing high-order mixed derivatives in physics-informed
neural networks using multi-index Bell polynomials}

\author{Fumihiro Imoto\corref{cor1}}
\ead{fumihiro.imo@gmail.com}
\cortext[cor1]{Corresponding author.}

\begin{abstract}
Physics-informed neural networks for high-order partial differential
equations require mixed input derivatives and their gradients with
respect to the network parameters. Standard implementations obtain an
order-$K$ derivative by repeated automatic differentiation. We instead
organize the forward recursion of the multivariate Fa\`a di Bruno
formula and its explicit backpropagation over a prescribed
downward-closed set of multi-indices, with the Bell-polynomial
convolutions tabulated once. The forward pass carries only the
derivatives the differential operator requires. The backward pass
propagates gradients from losses formed from any subset of them,
including nonlinear products and coupled fields.
Both recursions are exact up to roundoff and avoid nested computational
graphs. An independent Taylor-jet implementation, symbolic checks of
the test problems, and finite differences verify derivatives and loss
gradients through order seven. On one CPU
core, the method evaluates 330 mixed derivatives with respect to four
inputs through order seven and the corresponding loss gradient without
the memory failures observed for several nested implementations. Numerical tests
include third-, fifth-, and seventh-order dispersive equations,
incompressible flow, and a manufactured five-field electrohydrodynamic
system. The
seventh-order Zakharov--Kuznetsov test in $3+1$ dimensions has a
relative solution error of $6\times10^{-4}$.
\end{abstract}


\end{frontmatter}

%=====================================================================
\section{Introduction}
Physics-informed neural networks (PINNs) represent a field
$u_\theta(x)$ and determine its parameters from data, boundary
conditions, and differential-equation residuals
\cite{raissi2019,karniadakis2021}. These residuals require derivatives
of the network with respect to its inputs. First and second derivatives
are common. Higher derivatives are needed for the Kawahara equation
\cite{kawahara1972}, the Zakharov--Kuznetsov (ZK) equation
\cite{zk1974,infeld1985}, and related dispersive equations. Their
third-, fifth-, and seventh-order operators may also mix several
spatial variables and time.

Lagaris et al.\ \cite{lagaris1998} minimized differential-equation
residuals using trial functions that satisfy the boundary conditions.
Modern PINNs often include boundary conditions as penalty terms
\cite{raissi2019,karniadakis2021}, as implemented in packages such as
DeepXDE \cite{lu2021deepxde}. High-order PINNs remain difficult to
train. Sources of error include imbalanced loss terms
\cite{wang2021gradient}, optimization failures even for simple
equations \cite{krishnapriyan2021}, and poor conditioning
\cite{wang2022ntk}. Domain decomposition \cite{jagtap2020} and
gradient-enhanced residuals \cite{yu2022gradient} address some of these
problems. Chen et al.\ \cite{chen2024} also found that the activation
function affects fifth-order KdV calculations.

This work concerns the cost of the required derivatives. A standard
implementation applies forward- or reverse-mode automatic
differentiation (AD) repeatedly to obtain an order-$K$ derivative
\cite{griewank2008}. The resulting expressions and computational graphs
grow rapidly with $K$. Taylor-mode AD \cite{bettencourt2019} efficiently
computes directional derivatives, and separable architectures favor
forward-mode AD \cite{cho2023}. Chickering \cite{chickering2024} used
the univariate Fa\`a di Bruno formula to propagate derivatives of
arbitrary order during the forward pass, with the parameter gradient
obtained by framework backpropagation through that pass. Shi et al.\
\cite{shi2024stde} contract derivative tensors of arbitrary order with
Taylor-mode automatic differentiation and randomized jets, which
amortizes a differential operator over the optimization and scales to
very high input dimension; tailored jets recover selected mixed
partials as well. The present method differs in three respects: it is
deterministic, it returns a prescribed downward-closed family of mixed
partials in one pass rather than a contraction, and its weight gradient
is a closed-form reverse sweep rather than differentiation through a
forward-mode pass. The JAX jet column of Table~\ref{tab:speed} is the
cost of one Taylor-mode jet; recovering a full mixed set that way needs
one jet per direction of a polarization set. An operator such as
$\partial_x\nabla^{2j}$ requires several such derivatives, and PINN
training requires the gradient of the resulting loss with respect to
every network parameter.

A fully connected network consists of affine maps and elementwise
activations. The multivariate Fa\`a di Bruno formula gives its
high-order input derivatives in terms of partial Bell polynomials
\cite{constantine1996}. Backpropagation for losses that contain input
derivatives of a feed-forward network is not new: Avrutskiy
\cite{avrutskiy2017} derived forward and backward propagation for
arbitrary derivatives in any number of inputs, and Berg and Nystr\"om
\cite{berg2018} used the low-order case to solve PDEs in complex
geometries. Rodini \cite{rodini2022} gave the recursions through
second order, and our earlier work used a first-order form and its
adjoint for derivative-constrained density-functional learning
\cite{imoto2021}. What this work adds is the organization of these
recursions over a prescribed downward-closed multi-index set, with the
Bell-polynomial convolutions enumerated and cached once, a residual
seed for multicomponent systems and products of up to four derivative
factors, and a verified implementation. The implementation supports
orders through 15.

The method has three parts. A forward pass computes the required
mixed derivatives. The explicit backpropagation recursion gives the
parameter gradient of any loss formed from them. An automatically
generated downward closure omits derivatives that do not enter the
residual. We
verify the recursions independently, measure their cost against nested
AD, and apply them to scalar dispersive equations and coupled flow
systems.

Section~\ref{sec:method} derives the method and its computational cost.
Section~\ref{sec:implementation} describes the implementation and
verification tests. Section~\ref{sec:cost} reports performance, and
Section~\ref{sec:bench} gives numerical examples. Section~\ref{sec:activation}
examines the activation derivatives required at high order. The final
section summarizes the scope and limitations of the method.

%=====================================================================
\section{High-order network differentiation}
\label{sec:method}

\subsection{Setting and notation}
Let $u_\theta:\mathbb{R}^{D_0}\to\mathbb{R}^{D_N}$ be a fully
connected network with layers $l=0,\dots,N$, widths $D_l$, weights
$W^{(l)}$, and biases $b^{(l)}$.
Hidden layers use a $C^{K+1}$ elementwise activation $\sigma$; the
output is linear. Any such activation with known derivatives can be
used. The code provides $\tanh$, $\sin$, $\mathrm{erf}$, $J_0$, and
$J_1$; examples use $\tanh$ unless noted. The input
$\xi=(\xi_1,\dots,\xi_{D_0})$ contains all independent variables, such as
$(t,x,y,z)$. The recursions do not distinguish time from space. The
layer map is
\begin{equation}
\begin{aligned}
T^{(0)} &= \xi, &\qquad
S^{(l)} &= W^{(l)} T^{(l-1)} + b^{(l)}, \\
T^{(l)} &= \sigma\big(S^{(l)}\big) \ (l < N), &\qquad
T^{(N)} &= S^{(N)} = u_\theta,
\end{aligned}
\label{eq:layermap}
\end{equation}
where $T^{(l)},S^{(l)}\in\mathbb R^{D_l}$,
$u_{\theta,i}=T^{(N)}_i$, and $D_N=1$ gives a scalar field. For the
multi-index $\boldsymbol{\alpha}=(\alpha_1,\dots,\alpha_{D_0})$, define
$|\boldsymbol{\alpha}|=\sum_v\alpha_v$,
$\boldsymbol{\alpha}!=\prod_v\alpha_v!$, and
$\binom{\boldsymbol{\alpha}}{\boldsymbol{\beta}}=\prod_v
\binom{\alpha_v}{\beta_v}$. Let
\begin{equation}
T^{(l,\boldsymbol{\alpha})}_j = \partial^{\boldsymbol{\alpha}} T^{(l)}_j(\xi), \qquad
S^{(l,\boldsymbol{\alpha})}_j = \partial^{\boldsymbol{\alpha}} S^{(l)}_j(\xi)
\end{equation}
at a fixed input. The output derivatives are
$T^{(N,\boldsymbol{\alpha})}_i=\partial^{\boldsymbol{\alpha}}u_{\theta,i}$. Let
$\mathcal A$ be a finite, downward-closed index set containing $\boldsymbol{0}$,
and set $K=\max_{\boldsymbol{\alpha}\in\mathcal A}|\boldsymbol{\alpha}|$.

\subsection{Forward recursion}
The affine map gives
\begin{equation}
S^{(l,\boldsymbol{\alpha})} = W^{(l)} T^{(l-1,\boldsymbol{\alpha})}
  + b^{(l)}\,\delta_{\boldsymbol{\alpha},\boldsymbol{0}}.
\label{eq:fwdlin}
\end{equation}
For the elementwise activation, the multivariate Fa\`a di Bruno
formula \cite{constantine1996} gives, with
$a(\xi)=S^{(l)}_j(\xi)$,
\begin{equation}
T^{(l,\boldsymbol{\alpha})}_j = \sum_{q=0}^{|\boldsymbol{\alpha}|}
\sigma^{(q)}\big(S^{(l,\boldsymbol{0})}_j\big)\,
B_{\boldsymbol{\alpha},q}\big(\{S^{(l,\boldsymbol{\beta})}_j\}_{\boldsymbol{0} \prec \boldsymbol{\beta} \preceq \boldsymbol{\alpha}}\big).
\label{eq:faa}
\end{equation}
The partial Bell polynomial sums over unordered partitions of
$\boldsymbol{\alpha}$ into $q$ nonzero multi-indices:
\begin{equation}
B_{\boldsymbol{\alpha},q}(\{s_{\boldsymbol{\beta}}\}) =
\sum_{\substack{\{\boldsymbol{\beta}_1^{m_1},\dots,\boldsymbol{\beta}_r^{m_r}\} \\
\sum_s m_s \boldsymbol{\beta}_s = \boldsymbol{\alpha},\ \sum_s m_s = q}}
\boldsymbol{\alpha}! \prod_{s=1}^{r}
\frac{s_{\boldsymbol{\beta}_s}^{m_s}}{m_s!\,(\boldsymbol{\beta}_s!)^{m_s}},
\label{eq:bell}
\end{equation}
where $B_{\boldsymbol{0},0}=1$ and $B_{\boldsymbol{\alpha},0}=0$ for
$\boldsymbol{\alpha}\succ\boldsymbol{0}$. The input seeds are
$T^{(0,\boldsymbol{0})}=\xi$, $T^{(0,\boldsymbol{e}_v)}_k=\delta_{vk}$, and zero at
higher orders. Equations~(\ref{eq:fwdlin})--(\ref{eq:faa}) propagate
all derivatives in $\mathcal A$ in one pass. Partitions and
coefficients depend only on $(\mathcal A,q)$ and are tabulated once.
Appending a constant unit with value 1 and zero derivatives
incorporates biases into the matrix product.

For $\sigma=\tanh$, the derivatives follow from
$\sigma^{(q)}(a)=P_q(t)$, where $t=\tanh a$, $P_0=t$, and
$P_{q+1}=(1-t^2)P_q'$. The backpropagation recursion requires orders
through $K+1$.

\subsection{Backpropagation recursion}
\label{sec:backpropagation}
Let a differentiable scalar loss $L$ depend on any output derivatives
$T^{(N,\boldsymbol{\alpha})}_i$, including derivatives of coupled fields, and
write the adjoint as $\bar X=\partial L/\partial X$. The backward sweep follows
(\ref{eq:fwdlin})--(\ref{eq:faa}). Differentiating (\ref{eq:bell})
gives
\begin{equation}
\frac{\partial B_{\boldsymbol{\alpha},q}}{\partial s_{\boldsymbol{\beta}}} =
\binom{\boldsymbol{\alpha}}{\boldsymbol{\beta}} B_{\boldsymbol{\alpha}-\boldsymbol{\beta},\,q-1},
\qquad \boldsymbol{0} \prec \boldsymbol{\beta} \preceq \boldsymbol{\alpha}, \quad q \ge 1.
\label{eq:bellderiv}
\end{equation}
Thus no new Bell polynomials are evaluated during the backward sweep.
For each hidden neuron,
\begin{align}
\bar S^{(l,\boldsymbol{0})}_j &= \sum_{\boldsymbol{\alpha}\in\mathcal{A}}
  \bar T^{(l,\boldsymbol{\alpha})}_j \sum_{q=0}^{|\boldsymbol{\alpha}|}
  \sigma^{(q+1)}\big(S^{(l,\boldsymbol{0})}_j\big) B_{\boldsymbol{\alpha},q},
\label{eq:adj0}\\
\bar S^{(l,\boldsymbol{\beta})}_j &= \sum_{\boldsymbol{\alpha} \succeq \boldsymbol{\beta}}
  \bar T^{(l,\boldsymbol{\alpha})}_j \binom{\boldsymbol{\alpha}}{\boldsymbol{\beta}}
  \sum_{q=1}^{|\boldsymbol{\alpha}|}
  \sigma^{(q)}\big(S^{(l,\boldsymbol{0})}_j\big) B_{\boldsymbol{\alpha}-\boldsymbol{\beta},\,q-1}
  \quad (\boldsymbol{\beta} \succ \boldsymbol{0}),
\label{eq:adjb}\\
\bar T^{(l-1,\boldsymbol{\alpha})} &= W^{(l)\mathrm{T}} \bar S^{(l,\boldsymbol{\alpha})},
\qquad
\frac{\partial L}{\partial W^{(l)}_{jk}} =
\sum_{\boldsymbol{\alpha}\in\mathcal{A}} \bar S^{(l,\boldsymbol{\alpha})}_j\,
T^{(l-1,\boldsymbol{\alpha})}_k.
\label{eq:adjw}
\end{align}
The bias gradient is
$\partial L/\partial b^{(l)}_j=\bar S^{(l,\boldsymbol{0})}_j$. Losses depending
on $u$ require the $q=0$ term
$\sigma'(a)\bar T^{(l,\boldsymbol{0})}$ in (\ref{eq:adj0}). The loss enters only
through $\bar T^{(N,\boldsymbol{\alpha})}_i=\partial L/
\partial T^{(N,\boldsymbol{\alpha})}_i$. The identity output gives
$\bar S^{(N,\boldsymbol{\alpha})}=\bar T^{(N,\boldsymbol{\alpha})}$, so
(\ref{eq:adjw}) starts at layer $N$. One sweep covers all $D_N$ fields;
a unit seed gives one output derivative's gradient with respect to
$\theta$ (Section~\ref{sec:optimizers}).

\subsection{Residual seeding and the multi-index set}
\label{sec:closure}
First consider the scalar dispersive residual
\begin{equation}
R[u] = \sum_k c_k\, u^{\delta_k}\, \partial^{\boldsymbol{\alpha}_k} u,
\qquad \delta_k \in \{0,1\},
\label{eq:residual}
\end{equation}
which includes linear terms and advective terms of the form
$u\,\partial^{\boldsymbol{\alpha}}u$. Expanding
$\nabla^{2j}=(\sum_v\partial_v^2)^j$ also gives
$\partial_x\nabla^{2j}$. Section~\ref{sec:systems} treats more general
products. For
$L=\frac{\lambda_r}{2}R^2$, the seeds are
$\bar T^{(N,\boldsymbol{\alpha})}_i
 = \lambda_r R\, \partial R/\partial u_{i,\boldsymbol{\alpha}}$,
and vanish otherwise. Initial and boundary data seed only
$\boldsymbol{\alpha}=\boldsymbol{0}$. The required downward closure is
$\mathcal{A}=\{\boldsymbol{\gamma}:\boldsymbol{\gamma}\preceq\boldsymbol{\alpha}_k
\text{ for some } k\}$, constructed from the residual. The
seventh-order extended ZK operator needs 89 indices in four variables,
versus 330 in the dense set $|\boldsymbol{\alpha}|\le7$
(Table~\ref{tab:bench}). The closure contains every recursive
dependency, so dense and restricted runs are identical
(Section~\ref{sec:closurecost}).

\subsection{Products of derivative factors}
\label{sec:systems}
Coupled residuals require products absent from (\ref{eq:residual}); the
multi-output forward and backpropagation recursions remain unchanged.
Each term is
\begin{equation}
c_k \prod_{r=1}^{M_k} \partial^{\boldsymbol{\beta}_r} u_{i_r},
\qquad M_k \le 4,
\label{eq:systerm}
\end{equation}
where $u_i$ is field $i$. One factor is linear; two cover advection,
electric body force, and flux products; three cover compressible momentum
flux; and four give the quartic term in Section~\ref{sec:lax}. The code
limits $M_k$ to 4, although the product rule does not. Factor $r$ is
seeded by the product of all other factors; repeated factors add the
correct multiplicity. Tests cover product differentiation and multiple
outputs.

\subsection{Complexity}
For each input point and layer, (\ref{eq:fwdlin}) costs
$O(|\mathcal A|D_lD_{l-1})$. Direct enumeration of
(\ref{eq:bell}) costs
$\sum_{\boldsymbol{\alpha}\in\mathcal A}\sum_q q|B_{\boldsymbol{\alpha},q}|$
multiplications per neuron, where $|B_{\boldsymbol{\alpha},q}|$ denotes the
number of partition terms, counting the $q$ factors of each term and its
multiplication by $\sigma^{(q)}$. This count is $6.1\times10^4$ for dense
$D_0=4$, $K=7$, and $6.3\times10^3$ for the ZK closure. Instead, we use
\begin{equation}
B_{\boldsymbol{\alpha},q} = \frac{1}{q}
\sum_{\boldsymbol{0} \prec \boldsymbol{\beta} \preceq \boldsymbol{\alpha}}
\binom{\boldsymbol{\alpha}}{\boldsymbol{\beta}}\, S^{(l,\boldsymbol{\beta})}_j\,
B_{\boldsymbol{\alpha}-\boldsymbol{\beta},\,q-1},
\qquad B_{\boldsymbol{\alpha},1} = S^{(l,\boldsymbol{\alpha})}_j \ (\boldsymbol{\alpha} \succ \boldsymbol{0}),
\quad B_{\boldsymbol{0},1} = 0.
\label{eq:convolution}
\end{equation}
It follows from the generating identity
\[
\sum_{\boldsymbol{\alpha}} B_{\boldsymbol{\alpha},q}\,
\frac{\zeta^{\boldsymbol{\alpha}}}{\boldsymbol{\alpha}!}
= \frac{F(\zeta)^q}{q!}, \qquad
F(\zeta) = \sum_{\boldsymbol{\beta}\succ\boldsymbol{0}}
S^{(l,\boldsymbol{\beta})}_j\frac{\zeta^{\boldsymbol{\beta}}}{\boldsymbol{\beta}!}.
\]
Ordering indices by
degree makes every right-hand term available; zero terms are removed
in advance. The Bell cost falls to $3.6\times10^4$ multiplications per
neuron for dense $K=7$ and $4.7\times10^3$ for the ZK closure. Rather
than retain each Bell value, the forward pass stores
$h^{(l,\boldsymbol{\alpha})}_j=\sum_q\sigma^{(q+1)}(a)B_{\boldsymbol{\alpha},q}$ at one
extra multiply-add per sum. Then
$\bar S^{(l,\boldsymbol{\beta})}_j = \sum_{\boldsymbol{\alpha}\succeq\boldsymbol{\beta}}
\bar T^{(l,\boldsymbol{\alpha})}_j \binom{\boldsymbol{\alpha}}{\boldsymbol{\beta}}
h^{(l,\boldsymbol{\alpha}-\boldsymbol{\beta})}_j$ because
$B_{\boldsymbol{\gamma},q}=0$ for $q>|\boldsymbol{\gamma}|$. Valid
$(\boldsymbol{\alpha},\boldsymbol{\beta})$ pairs are stored in flat lists. Workspace is
$O(D_{\max}|\mathcal A|(N+K))$ words plus these tables, with
$|\mathcal A|=\binom{D_0+K}{K}$ for a dense set.

%=====================================================================
\section{Numerical implementation and verification}
\label{sec:implementation}
DNNF90 is written in standard Fortran 2008; the only feature beyond
Fortran 2003 is the \texttt{erf} intrinsic, and all other modules
except the imported Bessel routines pass a strict Fortran 2003 check.
The core uses allocatable
components but no pointers. An optional \texttt{ISO\_C\_BINDING}
interface supports calls from C, C++, and Python. BLAS is optional.
Replacing pointer-based work arrays with contiguous allocatable arrays
reduced the measured gradient time by about one third. The source
distribution also contains the inputs, tests, reference results, and
post-processing programs used in this paper.

DNNF90 extends an earlier Fortran implementation for training on first
derivatives \cite{imoto2021,imoto2019thesis}. The numerical kernels and
the input-driven trainer are separate. One module constructs the
multi-index sets, Bell tables, activation derivatives, and downward
closures. A second module implements
(\ref{eq:fwdlin})--(\ref{eq:adjw}). Each thread supplies contiguous
workspace in the order used by the kernels. Unless stated otherwise,
the code is compiled with gfortran using \texttt{-O3} and all calculations
use double precision. Indirect table access dominates for narrow
networks, whereas matrix multiplication dominates for wider networks
(Section~\ref{sec:layered}).

We use three verification tests. First, all 330 derivatives for
$D_0=4$ and $K=7$ agree with an independent multivariate Taylor-jet
implementation, written without the library, to below
$4\times10^{-19}$; the exact solutions and coefficient tables of the
test problems are checked separately with symbolic algebra. Second,
central differences
verify first- and second-order input derivatives to
$5\times10^{-9}$--$1.2\times10^{-8}$. They verify gradients of the
collocation loss with respect to the weights to $1.4\times10^{-9}$,
including every seventh-order backpropagation path. Third, tests cover products
of one to four factors, multiple outputs, batched evaluation, and every
activation used below. The tests call the same kernels as the numerical
examples.

Unless stated otherwise, the numerical examples use a fixed random
seed. The distributed reproduction package generates every table and
figure from the corresponding input and output data.

%=====================================================================
\section{Computational cost and performance}
\label{sec:cost}

\subsection{Comparison with automatic differentiation}
\label{sec:layered}
Standard AD implementations obtain order $K$ by repeated
differentiation, while operator-overloading libraries nest scalar
types. The comparison includes CoDiPack \cite{sagebaum2019}, XAD
\cite{xad}, and ADOL-C \cite{griewank1996}. The present method tabulates
the chain rule for the alternating affine maps and elementwise
activations. The same tables are used for every layer, neuron, and
input point.

We measure the time for one gradient, with respect to all weights, of a
loss formed from squared input derivatives. The tests use a
4--8--8--1 network, 20 input points, double precision, and one CPU core.
Implementations in the same language use identical points and weights.
The Fortran tests use the same network and objective with independently
stored weights. Every derivative and parameter gradient is checked
before timing. The full mixed-derivative and directional tests are
different workloads and are reported separately.

\begin{table}[t]
\centering
\caption{Milliseconds per all-weight gradient over 20 points of a
4--8--8--1 network. The left group uses all
$\binom{4+K}{K}$ mixed partials through order $K$; the right uses one
$K$th directional derivative. Entries are medians of three interleaved
runs. Dashes denote memory exhaustion.}
\label{tab:speed}
\scriptsize\setlength{\tabcolsep}{3.5pt}%
\begin{tabular}{@{}rrrrrrrrrr@{}}
\toprule
& & \multicolumn{3}{c}{full mixed-partial set} &
\multicolumn{5}{c}{one directional derivative} \\
\cmidrule(lr){3-5}\cmidrule(l){6-10}
$K$ & derivatives & PyTorch & JAX & this work & JAX (jet) & CoDiPack & XAD & ADOL-C$^{\ddagger}$ & this work \\
\midrule
1 & 5   & 1.7 & 0.10 & 0.051 & 0.098 & 0.17  & 0.13 & 0.18 & 0.024 \\
2 & 15  & 6.6 & 0.41 & 0.123 & 0.130 & 1.8   & 0.29 & 0.28 & 0.032 \\
3 & 35  & 25  & 1.6  & 0.316 & 0.169 & 19    & 0.85 & 0.42 & 0.041 \\
4 & 70  & 101 & 7.9  & 0.799 & 0.205 & 141   & 3.3  & 0.59 & 0.053 \\
5 & 126 & --- & ---  & 1.94  & ---   & 963   & 15   & 0.74 & 0.067 \\
6 & 210 & --- & ---  & 4.22  & ---   & 4990  & ---  & 0.95 & 0.082 \\
7 & 330 & --- & ---  & 9.04  & ---   & 38000 & ---  & 1.15 & 0.100 \\
\bottomrule
\end{tabular}\\[2pt]
{\footnotesize $^\ddagger$ hand-written construction on the ADOL-C
tape; see the text.}
\end{table}

Absolute times depend on language, compiler, and implementation. The
increase with derivative order is less sensitive to these differences.
For the nested methods, the geometric-mean factor per order is
3.1--8.5: 3.9 for PyTorch, 4.3 for JAX, 3.1 for XAD, and 8.5 for
CoDiPack
(Table~\ref{tab:speed}). PyTorch and JAX exhaust the available memory
beyond $K=4$, and XAD does so beyond $K=5$. CoDiPack takes 38~s at
$K=7$. The time rises by factors of 1.4 per order for the hand-written
ADOL-C construction and 1.3 for the present recursion. At $K=7$, the
two directional tests take 1.2 and 0.10~ms, respectively. JAX Taylor
mode grows by a factor of 1.28 per order through $K=4$, compared with
1.30 for the present method, but its absolute time is about four times
larger in this test. At $K=7$, the full-set calculation takes 9.0~ms
for 330 mixed partials; the directional calculation takes 0.10~ms and
returns eight values.

The only high-order term in the two-variable KdV residual is
$u_{xxx}$. The corresponding times are 1.75~ms for PyTorch, 0.131~ms
for nested JAX, 0.120~ms for JAX Taylor mode
\cite{bettencourt2019}, and 0.085~ms for the present method. Taylor
mode treats one direction at a time and therefore requires several
expansions for the mixed partials in $\partial_x\nabla^{2j}$.

ADOL-C's \texttt{tensor\_eval} \cite{griewank2000} computes all 330 mixed partials at $K=7$
in 0.26~ms per point, compared with 0.45~ms here, but does not compute
their dependence on the network weights. The taped-weight test obtains
that gradient by expressing the present recursion in ADOL-C types. It
therefore checks the implementation overhead of the recursion rather
than a standard ADOL-C driver for this complete task.

\subsection{Effect of the multi-index restriction}
\label{sec:closurecost}
\begin{figure}[t]
\centering
\includegraphics[width=0.45\textwidth]{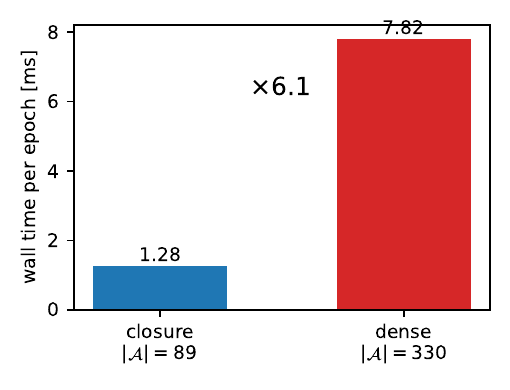}
\caption{Wall time per epoch for the seventh-order ZK benchmark using
all derivatives through order 7 or the downward-closed set required by
the residual. The two runs give identical training trajectories.}
\label{fig:closure}
\end{figure}

For seventh-order ZK, the automatically generated 89-index closure
takes 1.28~ms per epoch, compared with 7.82~ms for all 330 indices. The
6.1-fold difference is close to the ratio of table-operation counts.
All 121 final weights agree bitwise, so the restriction changes only
the intermediate calculations (Fig.~\ref{fig:closure}).

\section{Numerical tests and applications}
\label{sec:bench}

\subsection{Dispersive equations with exact solutions}

We use the amplitude-normalized hierarchy
\begin{equation}
\partial_t u + \nu\, u\, \partial_x u + b\, \partial_x \nabla^2 u
+ d\, \partial_x \nabla^4 u + g\, \partial_x \nabla^6 u = 0.
\label{eq:hierarchy}
\end{equation}
Here $\nabla^2$ is the spatial Laplacian, and $\nu$ absorbs amplitude
scaling. In $1+1$ dimensions, (\ref{eq:hierarchy}) gives KdV for
$d{=}g{=}0$, Kawahara for $g{=}0$ \cite{kawahara1972}, and a
seventh-order equation otherwise. In $3+1$ dimensions, it gives ZK
\cite{zk1974} and its higher-order extensions. The $1+1$ equations
have exact $\mathrm{sech}^{2m}$ solitons.
For $g=1$, $u=A\,\mathrm{sech}^6 k(x-ct)$ is a solution if
$A\nu=665280\,k^6$, $b=12304\,k^4$, $d=-200\,k^2$, and
$c=230400\,k^6$. Only $A\nu$ is fixed because rescaling $u$ changes
the nonlinear term. We set $A=1$ and hence $\nu=665280\,k^6$. A
plane-wave substitution gives a $3+1$-dimensional line soliton. If
$U(x-ct)$ solves the
one-dimensional equation with $(b_1,d_1,g_1)$, then
$u = U(\ell x + m y + n z - \ell c t)$ solves (\ref{eq:hierarchy})
with $b=b_1/L$, $d=d_1/L^2$, $g=g_1/L^3$,
$L=\ell^2+m^2+n^2$, and $\ell\ne0$, because
$u_t=-\ell c\,U'$, $u_x=\ell U'$, and
$\partial_x\nabla^{2j}u=\ell L^jU^{(2j+1)}$. The factor $\ell$
cancels, and we set it to one. An independent high-precision
calculation gives residuals below $10^{-30}$ for these solutions. They
cover two or four inputs and derivative orders 3, 5, and 7.

\subsection{Computational setup and results}
The two- and four-input cases use LeCun-initialized 2--16--16--1 and
4--8--8--1 $\tanh$ networks, respectively. Their data sets contain 100
and 180 exact initial or boundary values and 200 and 150 interior
points, respectively. Adam
\cite{kingma2015} minimizes weighted data and residual losses. The
exact solution is otherwise used only for error; the residual controls
early stopping and model selection.

\begin{figure}[t]
\centering
\includegraphics[width=\textwidth]{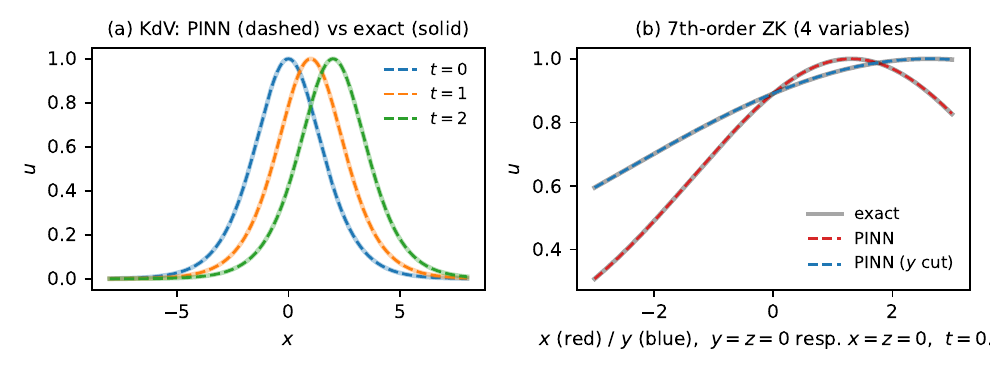}
\caption{Recovered solitary waves. (a) PINN (dashed) and exact (solid)
KdV solutions at three times. (b) $x$ and $y$ cuts through the
seventh-order extended ZK line soliton at $t=0.5$.}
\label{fig:solitons}
\end{figure}

\begin{figure}[t]
\centering
\includegraphics[width=\textwidth]{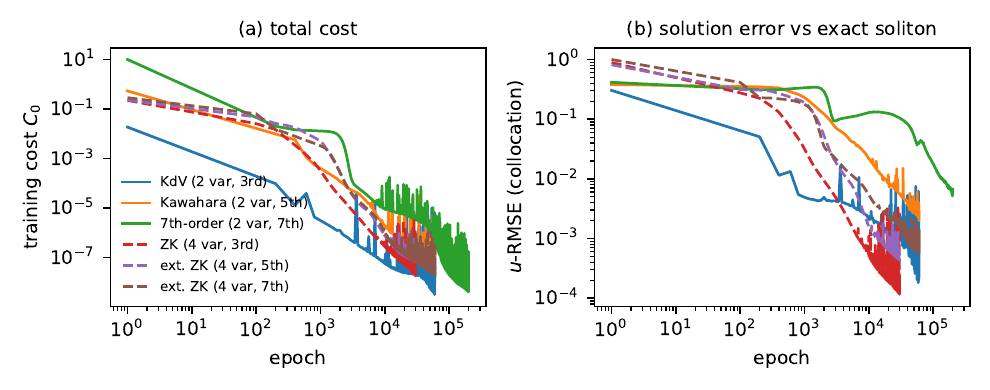}
\caption{Loss and exact-solution error for six benchmarks (solid:
$1+1$; dashed: $3+1$). Most descent occurs in the first few percent of
epochs; reported values are from the best epoch.}
\label{fig:histories}
\end{figure}

Table~\ref{tab:bench} and Figs.~\ref{fig:solitons}
and~\ref{fig:histories} give relative $u$-RMSEs of 0.02--0.59\% in $1+1$
dimensions and 0.01--0.06\% in $3+1$ dimensions. The errors generally
increase with differential order. The seventh-order ZK residual has 21
terms and requires 89 multi-indices; its relative $u$-RMSE is 0.06\%.
Each full-batch run trains for a fixed number of epochs; at the saved
best epoch, every residual loss reported here is below $10^{-7}$ per
point.

\begin{table}[t]
\centering
\caption{Fixed-seed benchmarks on one CPU core. ``Grad. error'' is
the largest sampled relative error against a central-difference
estimate. Loss per point and relative $u$-RMSE refer to the saved best
epoch.}
\label{tab:bench}
\footnotesize\setlength{\tabcolsep}{4.5pt}%
\begin{tabular}{@{}lccccccc@{}}
\toprule
equation & $D_0$ & order & $|\mathcal{A}|$ & grad.\ error & loss & rel.\ RMSE & time (s) \\
\midrule
KdV (2 var, 3rd)        & 2 & 3 & 5  & $8.1\times10^{-10}$ & $3.2\times10^{-9}$ & 0.0002 & 85 \\
Kawahara (2 var, 5th)   & 2 & 5 & 7  & $9.7\times10^{-10}$ & $7.1\times10^{-8}$ & 0.0024 & 114 \\
7th-order (2 var, 7th)  & 2 & 7 & 9  & $1.4\times10^{-9}$  & $4.1\times10^{-9}$ & 0.0059 & 664 \\
ZK (4 var, 3rd)         & 4 & 3 & 13 & $2.2\times10^{-10}$ & $1.8\times10^{-8}$ & 0.0001 & 99 \\
ext.\ ZK (4 var, 5th)   & 4 & 5 & 39 & $3.5\times10^{-10}$ & $7.1\times10^{-8}$ & 0.0004 & 231 \\
ext.\ ZK (4 var, 7th)   & 4 & 7 & 89 & $7.1\times10^{-10}$ & $1.9\times10^{-8}$ & 0.0006 & 1273 \\
\bottomrule
\end{tabular}
\end{table}

\subsection{Comparison with the Deep Ritz method}
\begin{figure}[t]
\centering
\includegraphics[width=\textwidth]{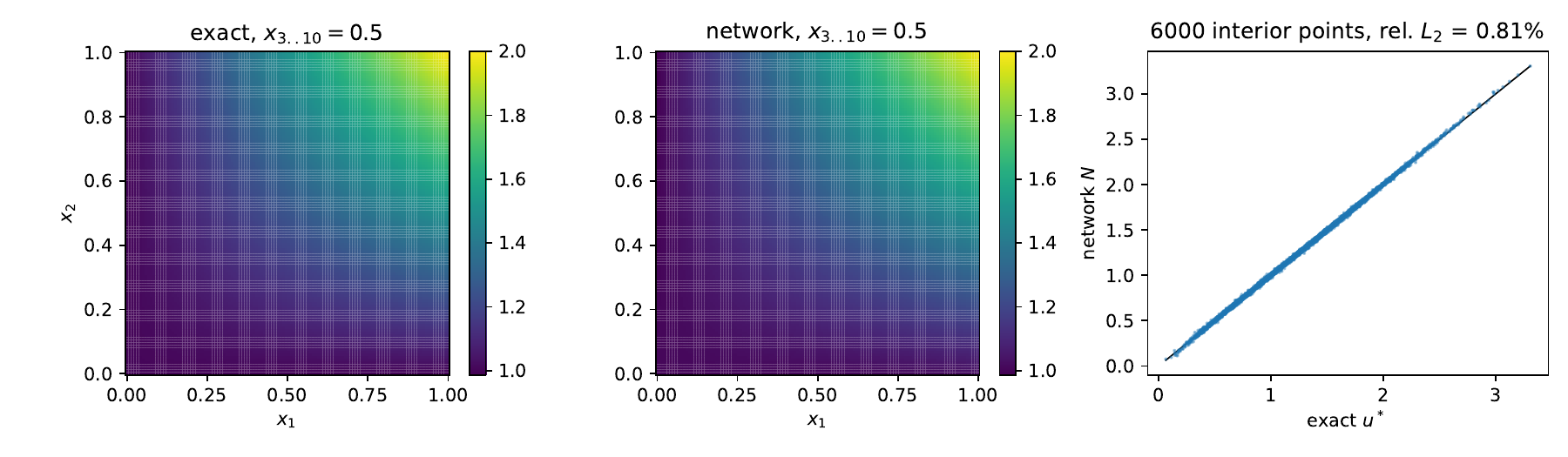}
\caption{Ten-dimensional Laplace problem (\ref{eq:eyu})
\cite{eyu2018}. Left: exact $(x_1,x_2)$ slice with other coordinates at
0.5. Middle: network on the same slice. Right: predictions against
exact values at 6{,}000 collocation points; relative $L_2$ error is
0.81\%.}
\label{fig:eyu}
\end{figure}

For comparison, we use E and Yu's Deep Ritz benchmark \cite{eyu2018}:
the Laplace equation on a ten-dimensional cube,
\begin{equation}
-\Delta u = 0 \ \text{on} \ (0,1)^{10}, \qquad
u = \textstyle\sum_{k=1}^{5} x_{2k-1} x_{2k} \ \text{on} \
\partial(0,1)^{10},
\label{eq:eyu}
\end{equation}
whose boundary value is the exact harmonic polynomial in
(\ref{eq:eyu}). Its ten-term residual needs only 21 indices. Training a
661-parameter 10--20--20--1 $\tanh$ network for 55,000 Adam epochs on
2000 boundary and 6000 collocation points takes 214~s on one core. The
relative $L_2$ error is 0.81\% on the collocation set and 0.82\% on
4000 new points (Fig.~\ref{fig:eyu}); E and Yu report 0.4\% with 671
parameters. The architectures and sampling procedures differ, so this
comparison concerns solution accuracy rather than execution time.

\subsection{Residual Jacobians and optimization}
\label{sec:optimizers}
Since the loss affects only the output seeds, a second backward pass
seeded by $\partial R/\partial u_{\boldsymbol{\alpha}}$ gives the exact residual
Jacobian $j_n=\partial R(\xi_n)/\partial\theta$. Damped Gauss--Newton
then requires one extra pass per point and one linear solve. The update
has the algebraic form of a natural-gradient step \cite{amari1998}, but
$G$ is the Gauss--Newton matrix of the residual loss. With
$G = \frac{1}{N_b}\sum_n j_n j_n^{\mathrm{T}}$, the step is
$\Delta\theta = -\eta\,(G + cI)^{-1}\nabla_\theta L$. We test absolute
damping, $c=\mu$, and relative damping,
$c=\mu\,\mathrm{tr}\,G$. The latter adjusts the damping to the scale of
$G$.

The same Jacobian can be used in an extended Kalman filter that
processes scalar observations by rank-one updates
\cite{blank1994}. Each collocation point supplies the residual $R$, its
target value zero, and the Jacobian $j_n$. For a nonlinear residual such
as $\nu uu_x$, the residual value and its derivative seed are evaluated
separately.

\begin{figure}[t]
\centering
\includegraphics[width=0.6\textwidth]{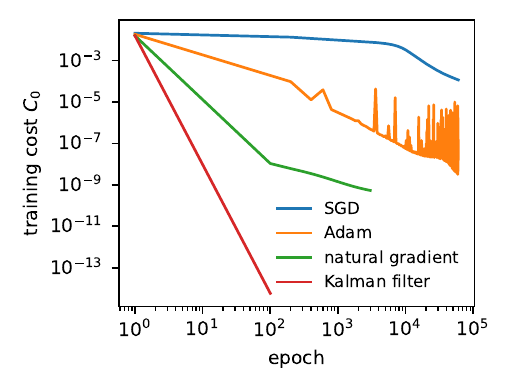}
\caption{KdV optimizer comparison with identical initial weights, data,
and 270-point full batches. Gauss--Newton and Kalman updates use exact
pointwise residual Jacobians. Results at equal wall time are given in
the text.}
\label{fig:optimizers}
\end{figure}

Figure~\ref{fig:optimizers} compares the optimizers on the 270-point
full-batch KdV problem. Damped Gauss--Newton reaches
$5.4\times10^{-10}$ in 3000 epochs; Adam and steepest descent reach
$3.2\times10^{-9}$ and $1.2\times10^{-4}$ in 60,000. Minibatch Adam
stalls near $2\times10^{-7}$ with 20 points; changing the learning rate
over the decade around the default, from $10^{-4}$ to
$3\times10^{-3}$, changes this by less than twofold.
Batch sizes of 20, 60, 135, and 270 lower the final loss
to $2.0\times10^{-7}$, $4.5\times10^{-8}$,
$3.0\times10^{-8}$, and $2.4\times10^{-8}$, indicating sampling noise.
The Kalman filter reaches $5.8\times10^{-15}$ in
100 epochs and a relative solution error of $4.2\times10^{-5}$. Its
270 rank-one updates cost 53~ms per epoch, versus 1.5~ms for steepest
descent and 28~ms for Gauss--Newton. At the same total time of 5.3~s,
steepest descent, Adam, and Gauss--Newton reach
$7.5\times10^{-3}$, $1.3\times10^{-6}$, and $6.2\times10^{-9}$.
The hierarchy and two-dimensional Poisson tests give the same ranking.
Quadratic covariance storage limits the filter to small networks; a
node-wise variant removes this limit, at a smaller initial covariance
and substantially more epochs per decade of loss reduction. All methods use the same
backpropagation recursion.
Section~\ref{sec:ehd} gives a different ranking.

\subsection{Seventh-order Lax equation}
\label{sec:lax}
\begin{figure}[t]
\centering
\includegraphics[width=\textwidth]{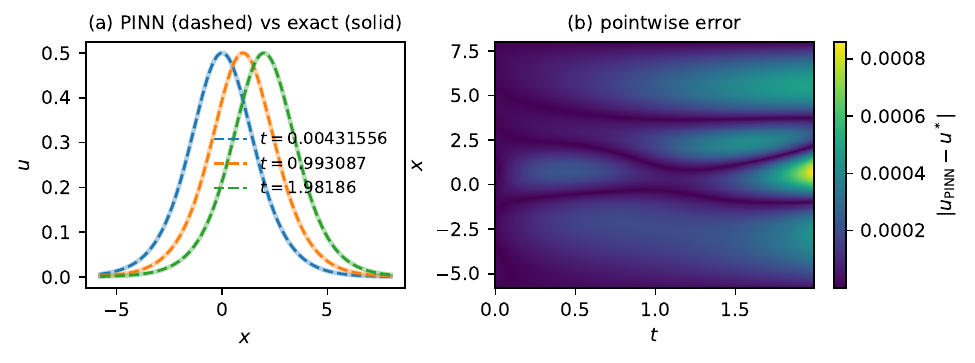}
\caption{Seventh-order Lax equation (\ref{eq:lax7}), whose term
$140\,u^3u_x$ contains four factors. (a) PINN (dashed) and exact
(solid) solitary waves at three times. (b) Pointwise error, with a
maximum of $8.6\times10^{-4}$.}
\label{fig:lax7}
\end{figure}

Lax's scalar seventh-order KdV equation \cite{lax1968} contains
products that (\ref{eq:residual}) cannot represent:
\begin{equation}
u_t + \partial_x\Big[35u^4 + 70\big(u^2u_{xx} + u\,u_x^2\big)
 + 7\big(2u\,u_{xxxx} + 3u_{xx}^2 + 4u_xu_{xxx}\big)
 + u_{xxxxxx}\Big] = 0.
\label{eq:lax7}
\end{equation}
The expansion has nine terms with one to four factors, including
$140u^3u_x$. For the exact wave
$u=\tfrac12\mathrm{sech}^2((x-t)/2)$, we train a 2--16--16--1 network
from random weights using full-batch L-BFGS, 100 data points, and 200
interior points. After 3000 epochs (39~s on one core), the
relative $L_2$ error is $8.8\times10^{-4}$ and $R^2=0.999999$
(Fig.~\ref{fig:lax7}). Chen et al.\ \cite{chen2024} report
$3.3\times10^{-3}$ for
$\tanh$ and $1.1\times10^{-3}$ for sine. The architectures, sampling,
and hardware differ, so these values show comparable accuracy rather
than a controlled performance comparison.

\subsection{Kovasznay flow}
\label{sec:kovasznay}
The Kovasznay flow \cite{kovasznay1948} is a steady exact solution of
the two-dimensional
incompressible Navier--Stokes equations,
\begin{equation}
u = 1 - e^{\lambda x}\cos 2\pi y, \qquad
v = \frac{\lambda}{2\pi} e^{\lambda x}\sin 2\pi y, \qquad
p = \frac{1}{2}\big(1 - e^{2\lambda x}\big),
\label{eq:kovasznay}
\end{equation}
with $\lambda=\dfrac{1}{2\nu}-
\sqrt{\dfrac{1}{4\nu^2}+4\pi^2}$. At $\mathrm{Re}=40$,
$\lambda=-0.96374054$. A 50-digit evaluation gives a residual of
$4\times10^{-51}$. This three-field, two-variable problem has three
residuals and 12 terms of the form (\ref{eq:systerm}).

\begin{figure}[t]
\centering
\includegraphics[width=\textwidth]{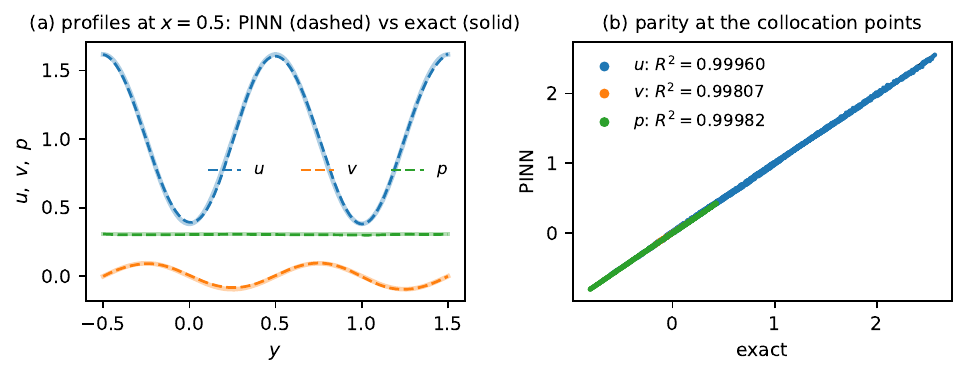}
\caption{Kovasznay flow at $\mathrm{Re}=40$. (a) PINN (dashed) and
exact (solid) fields along $x=0.5$. (b) Predictions against exact
values at collocation points, with $R^2$ for each field.}
\label{fig:kovasznay}
\end{figure}

A supervised fit gives $R^2=0.9992$, 0.9995, and 0.9986 for $u$, $v$,
and $p$. Another 400 full-batch L-BFGS epochs using boundary and
residual losses give 0.99960, 0.99807, and 0.99982
(Fig.~\ref{fig:kovasznay}).

Without boundary data, the residual falls by four orders of magnitude,
but $R^2$ becomes $-0.33$, 0.07, and $-0.16$. A small residual alone
does not identify the solution.

\subsection{Manufactured electrohydrodynamic system}
\label{sec:ehd}

The recursions also apply to a coupled model of electrostatics and
incompressible discharge flow:
\begin{align}
\nabla^2\phi + \rho/\varepsilon &= S_\phi, \nonumber\\
\mathbf{u}\!\cdot\!\nabla\rho
  - \mu\,\nabla\!\cdot\!(\rho\nabla\phi) - D\,\nabla^2\rho &= S_\rho,
\nonumber\\
\mathbf{u}\!\cdot\!\nabla\mathbf{u} + \nabla p - \nu\,\nabla^2\mathbf{u}
  + c\,\rho\nabla\phi &= \mathbf{S}, \nonumber\\
\nabla\!\cdot\!\mathbf{u} &= 0.
\label{eq:ehd}
\end{align}
This model has five fields, five residuals, and 25 terms. The electric
field transports charge and exerts a body force, while the velocity
advects charge. These couplings use (\ref{eq:systerm}); the drift term
also contains $\nabla\rho\cdot\nabla\phi+\rho\nabla^2\phi$. Each data
or residual term is divided by the squared scale of its component;
without this normalization, the largest field dominates the loss.

The optimizer ranking differs from the KdV result. Starting from a
supervised fit, fixed-step methods increase or barely reduce the
objective. Adam leaves the fitted region at $\eta=2\times10^{-3}$,
where the objective rises by an order of magnitude within one epoch. At
$2\times10^{-4}$ it holds the fitted region but after 3000 epochs
remains above $10^{-4}$, whereas L-BFGS reaches $1.2\times10^{-5}$ and
the Gauss--Newton updates below reach $1.1\times10^{-6}$.
Finite-difference directional checks agree
with the gradient to five digits, which indicates an optimization
rather than differentiation error. The sequential Kalman update leaves
the fitted solution while still reducing the residual: its forgetting
factor enlarges the covariance at every observation, and with thousands
of observations per epoch the gains do not decay. The residual falls by
four orders, but $R^2$ against the manufactured fields drops from 0.998
to $-23$ for the pressure, so the residual is being reduced along a
direction that leaves the solution. An error-based rejection test does
not prevent this failure and never triggers here. A forgetting factor
of one holds the solution, with every component above $R^2=0.97$, but
successive linearizations still produce oscillatory updates. Full-batch
L-BFGS avoids this sequential conflict.

\begin{figure}[t]
\centering
\includegraphics[width=\textwidth]{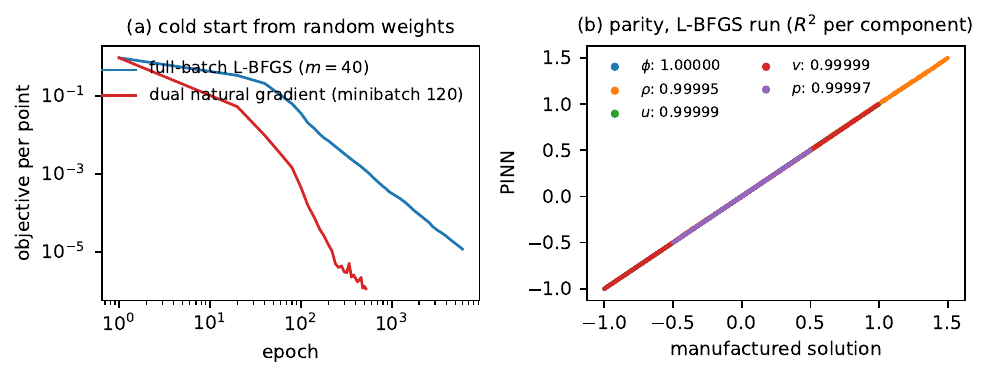}
\caption{Five-field electrohydrodynamics from random weights. (a)
Weighted objective for 40-pair full-batch L-BFGS and 120-point dual
Gauss--Newton. (b) L-BFGS predictions against the manufactured
solution, with $R^2$ for each field.}
\label{fig:ehd}
\end{figure}

L-BFGS with Armijo backtracking needs no prescribed learning rate. From
random weights, 40 curvature pairs lower the objective per point from
0.974 to $1.2\times10^{-5}$ in 6000 epochs (1055~s on one core).
The $R^2$ values for $\phi$, $\rho$, $u$, $v$, and $p$ are 1.00000,
0.99995, 0.99999, 0.99999, and 0.99997
(Fig.~\ref{fig:ehd}). At epoch 300, the losses are
$3.3\times10^{-3}$ with 40 pairs and $6.3\times10^{-3}$ with eight.

L-BFGS needs successive gradients of the same full-batch objective,
which can be expensive for large collocation sets. For minibatched
Gauss--Newton updates,
the metric must retain one Gauss--Newton row per residual, including
its $\sqrt{w_r}$ weight; summing residuals into one row per point loses
their separate directions. The Gram-space identity
\begin{equation}
(G + cI)^{-1} b = \frac{1}{c}\Big[\,b
  - \frac{1}{N_b} J^{\mathrm{T}} K^{-1} J b\Big],
\qquad K = cI + \frac{1}{N_b} J J^{\mathrm{T}},
\qquad c > 0,
\label{eq:dual}
\end{equation}
with $G=J^{\mathrm T}J/N_b$ matches the dense step to all printed
digits and costs $O(N_{\mathrm{row}}^2n_w+N_{\mathrm{row}}^3)$ rather
than $O(n_w^3)$, where $N_{\mathrm{row}}=N_bn_{\mathrm{res}}$. It cuts an
epoch from 120 to 0.6~s on the 120-point batches used below. With
adaptive Levenberg--Marquardt damping, these batches lower the objective
per point from 0.970 to $1.12\times10^{-6}$ in 520 epochs, reaching at
epoch 220 the value that full-batch L-BFGS attains only after 6000
epochs, and continue to descend (Fig.~\ref{fig:ehd}). A 40-point batch stalls at
$2.2\times10^{-4}$ through 6000 epochs. Tripling the batch size lowers
this apparent sampling-noise floor and gives nearly 200-fold lower loss
in one-twelfth as many epochs, consistent with
Section~\ref{sec:optimizers}.

\begin{figure}[p]
\centering
\includegraphics[width=0.92\textwidth]{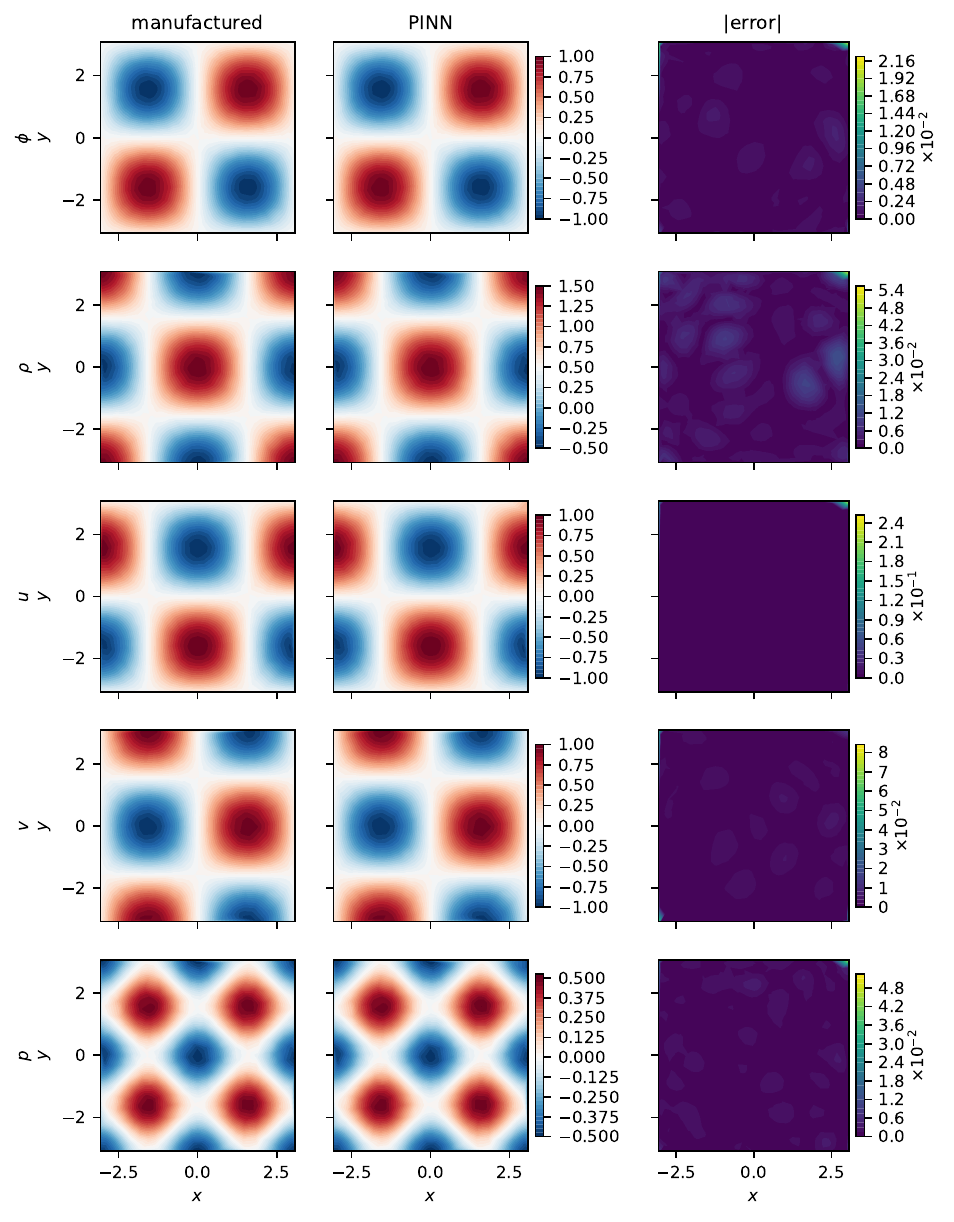}
\caption{Fields after L-BFGS training from random weights: manufactured
solution (left), network (middle), and error (right) for $\phi$,
$\rho$, $u$, $v$, and $p$. Panels interpolate scattered collocation
points; the four corners of the domain are added, with the
manufactured solution and the network each evaluated there, so that the
triangulation covers the square.
Labels give the error scales.}
\label{fig:ehdfields}
\end{figure}

Figure~\ref{fig:ehdfields} compares the manufactured and fitted fields.

Full-batch L-BFGS works when all points fit in one batch; dual
Gauss--Newton handles larger sets. Both use the same backpropagation
recursion.

%=====================================================================
\section{Effect of the activation function}
\label{sec:activation}
The backpropagation recursion uses $\sigma^{(q)}$ through $q=K+1$, so
$\sigma\in C^{K+1}$. Equation~(\ref{eq:faa}) multiplies these
derivatives by Bell polynomials. Their magnitude can therefore affect
both arithmetic scale and optimization. Network weights, depth, and
input scaling also matter.

The derivatives of $\tanh$ grow rapidly (Table~\ref{tab:sigma}), while
$|\sin^{(q)}(a)|\le1$. For Bessel functions,
\[
J_\nu^{(q)}(a) = \frac{(-1)^q}{2^q}
\sum_{k=0}^{q} (-1)^k \binom{q}{k} J_{\nu+q-2k}(a).
\]
Since $|J_n(a)|\le1$, this identity bounds every derivative of $J_0$
and $J_1$ by one \cite{dlmf10.6}. Their measured maxima decrease with
order.

\begin{table}[t]
\centering
\caption{Largest value of $|\sigma^{(q)}(a)|$ on $a \in [-6,6]$.}
\label{tab:sigma}
\begin{tabular}{@{}lrrrr@{}}
\toprule
$\sigma$ & $q=2$ & $q=4$ & $q=6$ & $q=8$ \\
\midrule
$\tanh$ & 0.77 & 4.1   & 52    & 1220 \\
erf     & 0.97 & 4.4   & 37    & 454 \\
$\sin$  & 1.0  & 1.0   & 1.0   & 1.0 \\
$J_0$   & 0.50 & 0.375 & 0.31  & 0.273 \\
$J_1$   & 0.41 & 0.335 & 0.289 & 0.257 \\
\bottomrule
\end{tabular}
\end{table}

Bessel-type functions have recently been tested as neural-network
activations \cite{martins2025,vieira2025,selmi2025}; sine has
outperformed $\tanh$ on fifth-order KdV equations \cite{chen2024}. We
therefore compare these activations in high-order derivative tests.

Loss weights can mask this effect. Taylor weights
$\lambda_p=1/(p!)^2$ weight order 7 by only $4\times10^{-8}$ relative
to order 1. Residuals constrain derivative combinations: seventh-order
ZK reaches 0.06\% solution error although its seventh derivatives are
twice the exact ones in norm over the ten seventh-order indices the
residual uses, and the worst of these exceeds its exact value by a
factor of 34 in maximum modulus. Over all 120 seventh-order indices,
most of which no residual term constrains, the norm ratio is 6.4.

To isolate the effect, we set $\lambda_p=0$ for $p<7$ and
$\lambda_7=1$:
\[
L = \frac{1}{2N} \sum_{n=1}^{N} \sum_{|\boldsymbol{\alpha}|=7}
\big(T^{\boldsymbol{\alpha}}(\xi_n) - y_{n,\boldsymbol{\alpha}}\big)^2.
\]
Values and lower derivatives remain unconstrained. All 330 indices
through order 7 are evaluated; 120 enter the loss.
Table~\ref{tab:seventh} reports eight paired runs with identical data,
architecture, schedule, and initial weights for each activation.
Hyperparameters were tuned for $\tanh$.

\begin{table}[t]
\centering
\caption{Training on seventh derivatives only, using eight paired
seeds. The relative error $\|T-y\|_7/\|y\|_7$ includes 120 derivatives
at all points; an identically zero prediction gives one.}
\label{tab:seventh}
\begin{tabular}{@{}lcccc@{}}
\toprule
$\sigma$ & loss median & loss range & rel.\ error & lower-loss pairs \\
\midrule
$\tanh$ & 0.224 & [0.208, 0.324] & 1.016 & --- \\
$\sin$  & 0.124 & [0.071, 0.196] & 0.753 & 8/8 \\
$J_0$   & 0.027 & [0.011, 0.063] & 0.357 & 8/8 \\
$J_1$   & 0.041 & [0.016, 0.077] & 0.442 & 8/8 \\
\bottomrule
\end{tabular}
\end{table}

Sine and both Bessel activations give a lower loss than $\tanh$ for all
eight paired seeds, and their loss ranges do not overlap. A one-sided
sign test gives $p=2^{-8}=0.004$ for each comparison and
Bonferroni-adjusted $p=0.012$. The relative error for $\tanh$ is near
one, which is no better in norm than a zero prediction. A 100-fold
learning-rate scan lowers its loss only to 0.193, still 2.5--17 times
the Bessel losses obtained at the default rate.
These eight pairs distinguish the three bounded derivative sequences
from $\tanh$ in this test, but do not rank sine, $J_0$, and $J_1$ among
themselves. The relative norm also does not separate shape and scale
errors.

The dispersive-wave collocation tests separate the activations only
when the seventh-order term is essential. Every residual term in the
benchmark hierarchy scales as
$k^7$ because
$\nu\sim k^6$, $b\sim k^4$, $d\sim k^2$, and
$\partial_x^ru\sim k^r$. Yet the seventh-order term is only 6\% of the
leading term in soliton $L_2$ norm, independent of $k$. The residual
can therefore remain small despite an inaccurate seventh derivative.
This helps explain why $\tanh$ remains accurate in these benchmarks.

A linear equation makes the highest spatial derivative indispensable:
\begin{equation}
u_t + u_{xxxxxxx} = 0.
\label{eq:linear7}
\end{equation}
Its exact solution $u=\cosh(\mu(x-\mu^6t))$ has seventh derivative
$\mu^7\sinh(\mu(x-\mu^6t))$, with $\mu^7=6.3$ at $\mu=1.3$. This sole
spatial term cancels $u_t$, so lower orders cannot absorb its error.
The solution is nonperiodic, so this test gives sine no direct
periodicity advantage.

\begin{table}[t]
\centering
\caption{Equation~(\ref{eq:linear7}) with
$u^* = \cosh(\mu(x-\mu^6 t))$, using five paired seeds. $R^2$ compares
the predictions with the exact solution at the collocation points.}
\label{tab:linear7}
\begin{tabular}{@{}lccc@{}}
\toprule
$\sigma$ & $R^2$ median & $R^2$ range & higher-$R^2$ pairs \\
\midrule
$\tanh$ & 0.690 & $[0.623, 0.815]$ & --- \\
$\sin$  & 0.915 & $[0.577, 0.966]$ & 3/5 \\
$J_0$   & 0.967 & $[0.953, 0.973]$ & 5/5 \\
$J_1$   & 0.940 & $[0.597, 0.952]$ & 4/5 \\
\bottomrule
\end{tabular}
\end{table}

Across five paired seeds on $[-2,2]\times[0,0.25]$, only $J_0$ gives a
higher $R^2$ than $\tanh$ in every pair (Table~\ref{tab:linear7}). Its
$R^2$ span is 0.020, compared with 0.192 for $\tanh$; sine and $J_1$
give higher values in three and four pairs, respectively.

Equation~(\ref{eq:linear7}) is used only as a manufactured collocation
test. The stated initial and boundary data do not define a well-posed
seventh-order initial-boundary-value problem. Table~\ref{tab:linear7}
therefore compares optimization under the specified loss; it does not
establish general convergence of a PDE solver.

Sine has a lower median residual than $J_0$ (0.58 versus 0.70) but a
less accurate solution. Residual and solution errors can therefore
rank models differently for operators with large null spaces.

At third order ($D_0=4$, $K=3$, 35 derivatives), sine and both Bessel
activations give a lower loss than $\tanh$ in all ten paired runs. A
one-sided sign test gives $p=2^{-10}=0.001$ for each comparison and
Bonferroni-adjusted $p=0.003$. In these runs the worst-derivative ranges
do not overlap (Table~\ref{tab:third}), and $\tanh$ gives negative worst
$R^2$ in four of them, below the mean-prediction baseline; the sign test
is reproducible, whereas the individual values follow the arithmetic
environment.

\begin{table}[t]
\centering
\caption{Third-order test with $D_0=4$, $K=3$, 35 derivatives, and ten
paired initializations. ``$R^2$ worst'' is the least-accurate derivative
in each run; ``min'' is its minimum over the ten seeds.}
\label{tab:third}
\begin{tabular}{@{}lcccc@{}}
\toprule
 & $R^2$ mean & $R^2$ worst & $R^2$ worst & loss \\
$\sigma$ & median & median & min & median \\
\midrule
$\tanh$ & 0.862 & 0.123 & $-0.761$ & $6.1\times10^{-3}$ \\
$\sin$  & 0.950 & 0.708 & 0.571    & $1.4\times10^{-3}$ \\
$J_0$   & 0.987 & 0.857 & 0.539    & $3.6\times10^{-4}$ \\
$J_1$   & 0.974 & 0.818 & 0.527    & $1.0\times10^{-3}$ \\
\bottomrule
\end{tabular}
\end{table}

The mean $R^2=0.86$ masks these failures. Although $\tanh$ gives field
$R^2=1.000$ in third- and fifth-order benchmarks, individual high
derivatives lose accuracy first. Their individual accuracy matters when
the derivatives themselves, rather than only the field, are quantities
of interest.

%=====================================================================
\section{Summary and limitations}
We organized the forward recursion and the explicit backpropagation
for high-order mixed derivatives of fully connected neural networks
over a prescribed downward-closed multi-index set. Multi-index Bell
polynomials propagate the input derivatives, and the backpropagation
recursion gives the gradient of a loss formed from them in one backward
pass. Restricting the calculation to the
residual's downward closure
reduces the seventh-order ZK test from
330 to 89 multi-indices without changing the weights or training
trajectory. An independent Taylor-jet implementation, symbolic checks
of the test problems, and finite differences verify the implementation
through order seven.

In the directional tests, the times for nested methods grow by
geometric-mean factors of 3.1--8.5 per derivative order, compared with
about 1.3 for the
present recursion. Several nested implementations exhaust the
available memory before order seven. The method also treats nonlinear
products and multiple fields. The tests cover third-, fifth-, and
seventh-order dispersive equations, the Kovasznay flow, and a
five-field electrohydrodynamic problem on one CPU core. The same
backpropagation recursion supplies the residual Jacobians used by
damped Gauss--Newton and Kalman updates. Two of these tests show the
same failure from different causes: the Kovasznay system without
boundary data, and the Kalman filter under its forgetting factor, both
reduce the residual by four orders while the solution moves away from
the exact one. Reported residuals therefore need an independent
accuracy measure.

The method is specialized to networks made of affine maps and smooth
elementwise activations; it does not replace general AD for arbitrary
computational graphs. It requires derivatives of the activation
through order $K+1$, and a dense mixed-derivative set still grows as
$\binom{D_0+K}{K}$. The reported timing comparisons are
implementation-dependent and use one CPU core. The differential-
equation examples are controlled tests with exact or manufactured
solutions, not new physical simulations.

Possible extensions include variable coefficients, transcendental
field nonlinearities, products of more than four factors, and
vectorized or parallel batches. Per-point Jacobians can also be used
for inverse problems and, with an explicit noise model, for Fisher
information.

\section*{Code and data availability}
The DNNF90 source code and the materials needed to reproduce the
reported results are available at \url{https://github.com/fimoto/DNNF90}. They include the
implementation, benchmark inputs, reference outputs, verification
tests, and programs used to generate the tables and figures. Each
setting quoted in the text has an input file beside the case it varies. The code
extends the version developed for F.~Imoto's thesis
\cite{imoto2019thesis}. No external data sets were used.

\section*{Acknowledgments}
The author received no external funding.
The author thanks M. Shimizu for discussions of numerical fluid
dynamics.

\section*{Declaration of competing interest}
The author declares no competing financial interests or personal
relationships that could have influenced this work.

\appendix

\section{Activation derivatives}
For $\sigma=\tanh$, $\sigma^{(q)}(a)=P_q(\tanh a)$, where
$P_0(t)=t$, $P_{q+1}(t)=(1-t^2)P_q'(t)$, and $\deg P_q=q+1$.
Equation~(\ref{eq:adj0}) requires orders through $K+1$. Any
$C^{K+1}$ elementwise activation with computable derivatives can be
used. Piecewise-linear activations cannot.

\section{Bell coefficients}
The coefficient for the partition
$\{\boldsymbol{\beta}_1^{m_1},\dots,\boldsymbol{\beta}_r^{m_r}\}$ is
$\boldsymbol{\alpha}!/\prod_s m_s!(\boldsymbol{\beta}_s!)^{m_s}$. The code enumerates
nondecreasing lists of parts once. Within the supported range, the
coefficients are integers stored in double precision. For $D_0=4$ and
$K=7$, all 16,779 dense terms agree with an independent enumeration.

\section*{Declaration of generative AI use in manuscript preparation}
During preparation of this work, the author used Claude Opus 5
(Anthropic) to refactor parts of the numerical code and ChatGPT
(OpenAI) to improve the language and readability of the manuscript.
The author reviewed and verified the resulting text and code and takes
full responsibility for the content of the article.

\end{document}